\input{aipcheck}

\documentclass[
    ,final            % use final for the camera ready runs
  ]
  {aipproc}

\layoutstyle{8x11double}

\begin{document}

\title{New Results of the Konkoly Blazhko Group}


\author{J. Jurcsik}{
  address={Konkoly Observatory, H-1525 Budapest, PO Box 67, Hungary}
}

\begin{abstract}
During the recent years the Konkoly Blazhko Group (PIs Johanna Jurcsik 
and B\'ela Szeidl, co-workers \'Ad\'am S\'odor, Zsombor Hurta and several 
undergraduate, graduate students) published new important results of Blazhko 
variables in 15 reviewed Journal articles. These results utilize 
multicolor CCD observations obtained with an automatic 60 cm telescope, 
and also previously unpublished Konkoly archive photometric data.
Our light curves are the most extended multicolor data-sets ever 
obtained for a Blazhko variable, the observations cover each phase of 
the pulsation and the modulation as well. We have detected many 
previously unknown features of the light curve modulation, and based on 
the different band's observations we also revealed the underlying 
variations of the mean physical parameters during the Blazhko cycle. 
In my contribution the main achievements of the Konkoly Blazhko Group 
are summarised.

\end{abstract}

\maketitle

%%%%%%%%%%%%%%%%%%%%%%%%%%%%%%%%%%%%%%%%%%%%
%% MAINMATTER
%%%%%%%%%%%%%%%%%%%%%%%%%%%%%%%%%%%%%%%%%%%%

\section{The Konkoly Blazhko Survey 2004-2009}

The aim of the Konkoly Blazhko Survey\footnote{http://www.konkoly.hu/24} is to obtain extended, accurate, multicolor light curves of modulated  and unmodulated RRab stars in order to 
\begin{itemize}
\item
 derive the incidence rate of the modulation in the sample \cite{kbs}; 
\item
 study long term changes in the pulsation and modulation properties \cite{rrg2,dmc}
\item 
find any changes in the physical parameters of the stars during the Blazhko cycle \cite{mw2,dmc,sodor}. 
\end{itemize}
The results utilize recent CCD observations obtained with an automatic 60 cm telescope (Fig. 1) and also archive Konkoly photometric data spanning more than 75 years.

\begin{figure*}
  \includegraphics[height=.45\textheight]{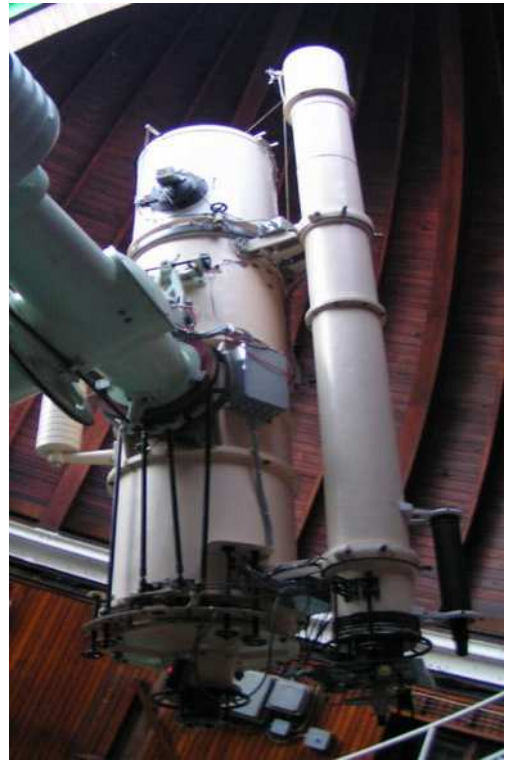}
  \caption{The automatized 24" telescope, Konkoly Observatory, Budapest, Sv\'abhegy.}
\end{figure*}

\section{Results}

\begin{figure*}
  \includegraphics[height=.53\textheight]{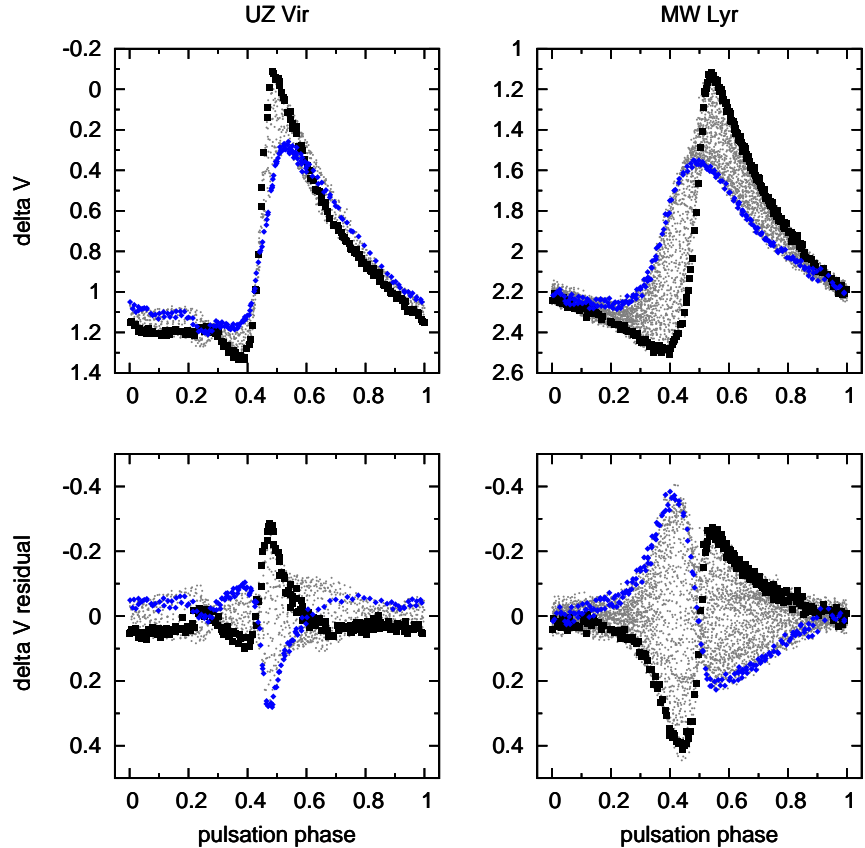}
  \caption{The observed light curves of UZ Vir and MW Lyr plotted according to the pulsation phase. Data belonging to the largest and smallest amplitude phases of the modulations are shown by different symbols. Bottom panes show the residual light curves of the corresponding data after the removal of the mean pulsation light variations.}
\end{figure*}

The observed light curves of Blazhko stars covered the whole pulsation cycle of the variables in different phases of the modulation, which made it possible to perform the most thorough analyses of the Blazhko type light variation that has never been published before (e.g., \citep{mw1}). In Fig. 2 the pulsation and modulation light variation (after the removal of the mean pulsation light curve) of MW Lyr and UZ Vir are plotted. The largest and smallest amplitude phases of the modulation are shown with different symbols. The amplitude of the modulation is the largest at around minimum-maximum phase of the pulsation, but in different stars the modulation pattern can be significantly different as shown in the figure.

As our data were the first extended, multicolor observations of Blazhko stars we could also show for the first time how the color-magnitude and two-color loops of the pulsation light and color variations changed in the different phases of the modulation  \citep{mw2}.

\begin{figure*}
  \includegraphics[height=.44\textheight]{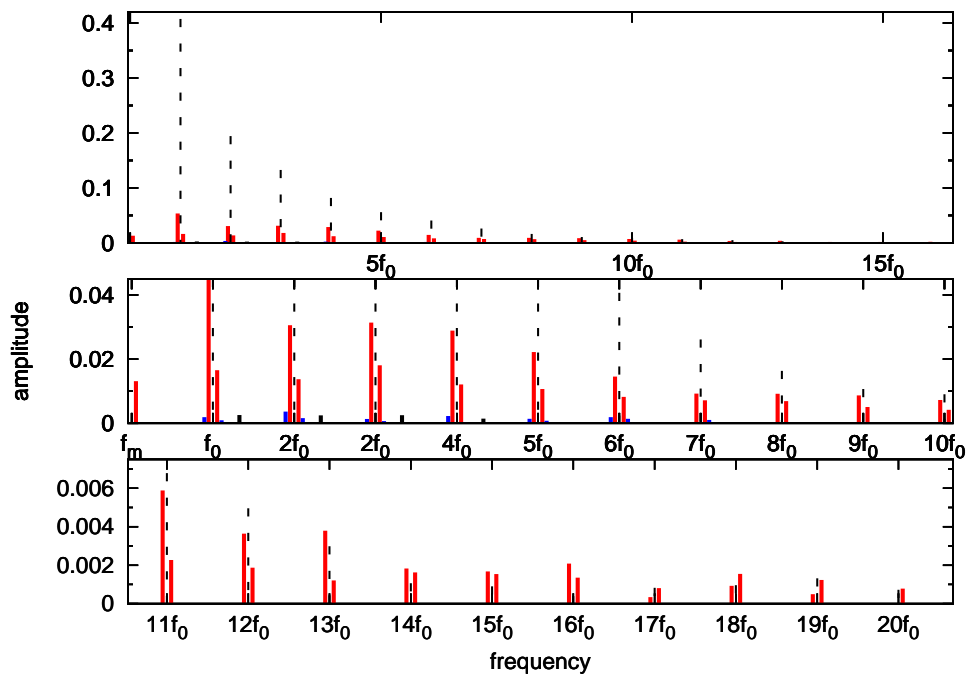}
  \caption{Schematic amplitude spectrum of UZ Virginis. The top panel shows the spectrum up to 16 $f_0$ frequency, indicating that while the pulsation frequencies have amplitudes with exponential-like decrease with increasing harmonic orders, the amplitude decrease of the modulation side frequencies is less steep, it is rather linear than exponential. The middle panel enlarges the modulation components up to 10 $f_0$. The modulation side frequencies are triplet components with highly asymmetric amplitudes ($f_0\pm f_m$), and small amplitude quintuplet components at $f_0\pm 2f_m$ frequencies. A secondary modulation frequency series also appears with large separation from the pulsation components. This series appears only at the larger frequency sides of the pulsation frequencies ($f_0 + f_m'$). The modulation frequency itself appears in the spectrum without question, too. The bottom panel, which shows the spectrum in the 10-20 $f_0$ frequency regime, documents that for a large modulation amplitude Blazhko star as UZ Vir is, the modulation frequency components can have larger amplitudes than the pulsation components at the higher harmonic orders. }
\end{figure*}

Our Blazhko light curves were the first time series suitable to study the 'fine structure' of the Fourier spectrum of the data. We found that the Fourier spectrum of Blazhko data can be characterized by the followings (Fig. 3 shows a schematic example).

\begin{itemize}
\item
Triplets, quintuplets and higher order multiplets appear around the pulsation frequency components which may have very asymmetric amplitudes \cite{rvuma,mw1,dmc}. The quintuplet and higher order multiplet components have only very small amplitudes. We have shown in \cite{szb} that the Fourier spectrum of an amplitude and phase modulated harmonic signal shares these properties.
\item
The modulation frequency, $f_m$, appears unambiguously in the spectra.
\item 
Secondary modulations which have either commensurable amplitudes to the main modulation components or very small amplitudes may also occur \cite{uzu,czl,mw1}.
\item
While the decrease of the amplitudes of the pulsation components with increasing orders is exponential-like, the decrease in the amplitudes of the modulation components is more linear with significant, random variations especially in the low order (1-3) components.
\item
The modulation side frequencies have similar (or even larger) amplitudes as the pulsation frequencies in the large harmonic orders.
\end{itemize}

Any reliable, valid explanation of the Blazhko effect has to explain all these properties of the light curve variations.

Blazhko RRab stars that were previously discovered showed modulation with relatively large amplitude (the maximum brightness variation was larger than $\sim 0.1$ mag) and with modulation period longer than 10 days. In the Konkoly Blazhko Survey, however, we found modulations with very small amplitude and with periods shorter than 10 days, too \cite{rrg1,ssc,dmc}. As the modulation side frequencies of these variables have only some mmag amplitudes they can be detected only in data of very low noise level. The short period of the modulation warns that in order not to miss short modulation period Blazhko stars in Survey data, frequency search for modulation side frequencies at least a $\pm 0.2$ cd$^{-1}$ vicinity of the pulsation frequency components has to be performed.

The Konkoly Blazhko Survey is the first systematic accurate survey that is dedicated to determine the frequency of the occurrence of the modulation in bright Galactic field RRab stars. Contrarily to previous estimates we found a  large percentage, about $ 50\%$, of RRab stars showing the Blazhko effect  \cite{kbs}, most probably due to the discovery of small amplitude modulations. 

Another important, and previously neglected, disregarded property of the Blazhko modulation is its temporal occurrence in some cases as we documented for RR Gem \cite{rrg2} and RY Com \cite{kbs}. The temporal behavior of the modulation warns that the Blazhko effect might be a universal property of RR Lyrae stars.

\subsection{Interpretation of the data} 

We have shown that the light curve modulation can be separated to dominantly amplitude and phase modulation components with a simple transformation corresponding to the phase variation of the main pulsation component ($f_0$) in different phases of the modulation as shown in Fig. 4 \cite{mw1,dmc}. Taking into account that, analytically, phase modulation is equivalent with periodic changes of the pulsation period, we interpret the phase variation of the main pulsation component as period changes of the fundamental mode oscillation. 

\begin{figure}
  \includegraphics[height=.34\textheight]{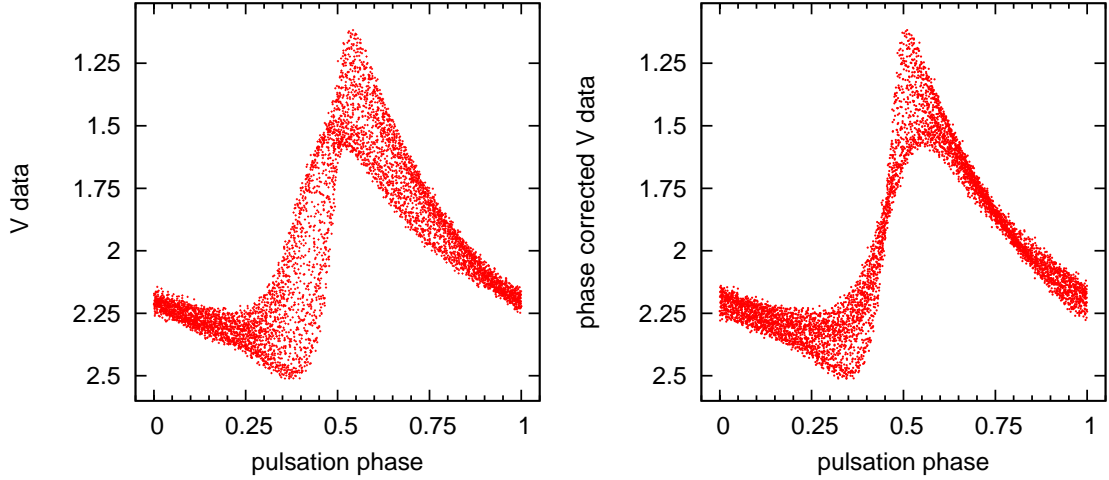}
  \caption{V light curve of MW Lyr. Left-hand panel:original data; right-hand panel: data corrected according to the phase variation of the main pulsation frequency, $f_0$.}
\end{figure}

\begin{figure}
  \includegraphics[height=.47\textheight]{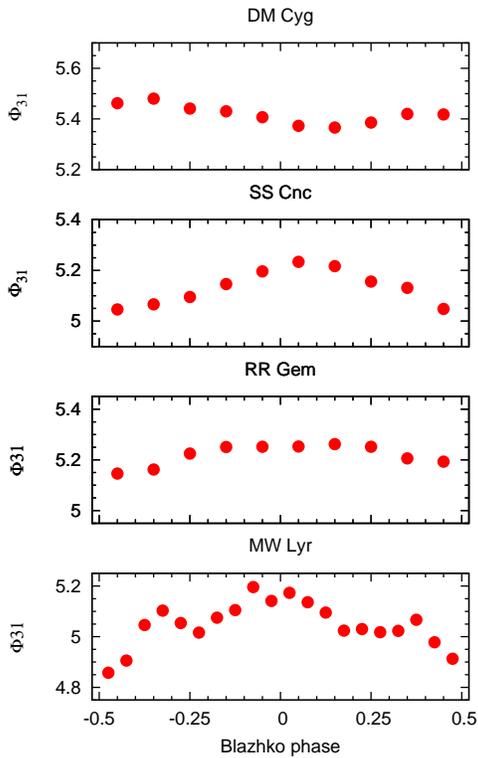}
  \caption{The variation of the $\Phi_{31}$ phase differences of the light curves of four Blazhko stars in our sample in different phases of the modulation. Though the variations of the phase differences are in the range of about only 0.1-0.3 rad, they clearly show that the phase modulations of the harmonic components of the pulsation are not perfectly coherent.}
\end{figure}

Analysing the light curves' shapes in different phases of the modulation shows that the phase-differences of the harmonic components ($\Phi_{k1}$) of the light curve do also show some minor changes, which means that the phase modulations of the pulsation components in different orders are not exactly the same (see Fig. 5 for the variations of $\Phi_{31}$ in three small and one large modulation amplitude Blazhko stars). Consequently, more precisely, when correcting the data for the phase variation of the main pulsation component, the residual variation of the light curves' shape is dominantly amplitude modulation but with small but nonnegligible changes also in the phase relations of the harmonic components.
This result proves that any description of the modulation that assumes that the  phase modulation is exactly the same in the different orders \cite{benko} cannot give an exact, accurate light curve solution.

Our observations and the interpretation of the results indicate that the physically meaningful frequency of the Blazhko effect is the modulation frequency, $f_m$, and not any of the modulation side frequencies. The following arguments support this conclusion:
\begin{itemize}
\item
the pulsation period and the amplitude (and shape) of the radial mode pulsation change with the modulation period;
\item
the mean global physical parameters of the star ($<R>,<L>,<T_{eff}>$) vary  $0.1-4\%$ during the Blazhko cycle \cite{mw2,dmc,sodor,ipm};
\item
$f_m$ has only small amplitude, but if the dominant frequency were e.g., $f'=f_0+f_m$, then, as a first order linear combination frequency, $f_m= f'-f_0$ should have to have  significantly larger amplitude than detected; 
\item
the amplitude and phase relations of the side frequencies in the different wavelength bands are very similar to the amplitude and phase relations of the pulsation components, while these properties of the modulation frequency are different, indicating that $f_m$ is the independent frequency characterizing the light curve variations \cite{ssc,mw1}.
\end{itemize}

We thus conclude that, most probably,  with the Blazhko cycle periodic, structural changes in the stars occur which result in period and amplitude (light curve) changes of the pulsation.

At present, there is no explanation of the Blazhko effect that would successfully explain all the observed properties of the modulation. Models which connect the modulation to the rotation of the star \cite{sh00} fail to explain multiple periodic modulations. The same is true for any resonance hypothesis \cite{kg}, the lack of a strict connection between the pulsation and modulation frequencies also contradicts the 2:1 resonance solution. Accounting additional nonradial modes, on the one hand, is not working theoretically \cite{dm}, and on the other hand would connect the dominant frequency of the Blazhko effect to a side-lobe frequency. Taking into account the above arguments this is most probably not the case. The most promising model of the light curve modulation seems to be the idea of Stothers \cite{stothers}, which presumes that with the modulation period the radial mode pulsation of the star varies due to a periodic turbulent convection dynamo acting in Blazhko stars. However, apart the lack of accurate modelling of such a phenomenon, this model would also meet difficulties to explain multiple modulations. Moreover, the strictly regular behavior of the modulation observed in many Blazhko stars would be also hard to originate from any magnetic dynamo effect. 

As final conclusion we can only say that further theoretical studies are needed to identify the true triggering mechanism of the Blazhko modulation.

\begin{theacknowledgments}
The financial support of the Hungarian OTKA grant T-068626 is acknowledged.
\end{theacknowledgments}

\bibliographystyle{aipproc}   % if natbib is available

\end{document}